\documentstyle[epsfig]{article}
\newcommand{\idn}{\hspace{\parindent}}
\newcommand{\be}{\begin{equation}}
\newcommand{\ee}{\end{equation}}
\newcommand{\ba}{\begin{eqnarray}}
\newcommand{\ea}{\end{eqnarray}}
\newcommand{\bb}{\begin{thebibliography}}
\newcommand{\eb}{\end{thebibliography}}
\newcommand{\MeV}{{\rm MeV}}

\renewcommand{\Im}{{\rm Im}}
\renewcommand{\Re}{{\rm Re}}

\begin{document}
\title{Pion and Kaon Electromagnetic Form Factors in
a $SU_{L}(3)\otimes SU_{R}(3)$ Effective Lagrangian}
\author{Yun Chang Shin$^{1)}$, Bong Soo Han $^{1)}$,
Myung Ki Cheoun$^{1,2)}$\thanks{Corresponding Author,
cheoun@phya.snu.ac.kr}, \\ K.S Kim$^{1)}$, Il-Tong Cheon $^{1)}$
\\
1) Department of Physics, Yonsei University, Seoul, 120-749,
\\
Korea
\\
2) IUCNSF, Seoul National University,151-742 Korea}

\date{15 July, 2000}
\maketitle
\begin{abstract}
A $SU(2)$ effective lagrangian is extended to a $SU_{L}(3)\otimes
SU_{R}(3)$ by including the vector and axial vector meson. With
this effective lagrangian, electromagnetic form factors of charged
pion and kaon are calculated in both time and space like regions.
The pseudoscalar meson loops are taken into account. Good
agreement with experimental data is obtained for those form
factors and charged pseudoscalar meson radii. Decay widths of
$\rho\rightarrow \pi\pi$ and $\phi\rightarrow K^{+}K^{-}$ are also
calculated and shown to agree with experimental data very well.
\end{abstract}
\section{Introduction}
\idn At energy below 1 GeV, the vector meson plays an important
role in electromagnetic interactions of the hadron. The vector
meson dominance model (VMD) has been proved remarkably successful
in the description of electromagnetic form-factors and decays of
the hadron, although it is a phenomenological approach. Many
approaches, such as a hidden gauge symmetry approach
(HGS)\cite{bando1}, a massive Yang-Mills approach
(MYM)\cite{meissner}, and so on, have been developed to include
the vector meson in a fundamental manner. By taking higher order
terms into account, redefining suitable field and adjusting
parameters, all of the model can be shown to be
equivalent\cite{birse}. However, a simple addition of higher order
terms is not a convenient method for those calculations. In our
previous paper\cite{park}, we have proposed an effective chiral
Lagrangian for the description of vector and axial-vector mesons
by considering all the relevant symmetries and the low-energy
constraints from chiral perturbation theory(ChPT). In that paper,
relevant experimental data are reproduced with only mass terms and
kinetic terms of spin-1 meson fields. The spin-1 mesons are
introduced in the non-linear realization of chiral symmetry, with
which it is easy to check consistency with chiral perturbation
theory. In constructing our model Lagrangian, we have stressed
simplicity. Only mass terms and kinetic terms of spin-1 meson
fields are necessary to meet experimental results. In some
approach, such as $O(p^{4})$ order expansion of ChPT, ${\cal
L}_{2}$ lagrangian gives loop contribution as well known\cite{gasser},
which
helps a good phenomenological description. But our effective
lagrangian theory, which is aimed for large energy process, uses
$O(p^2)$ expansion because most of the higher order contributions
in other approaches are incorporated by a single change in the
kinetic terms of vector field with only one parameter in our
model. Since full reviews concerning effective theories and their
relationships to other approaches can be found in other papers \cite{meissner,birse},
we skip them here.

In this paper our previous lagrangian, with a brief summary, is
extended to $SU(3)$ in section 2. In section 3, the
electromagnetic pion and kaon form factors and some related decays
with this lagrangian are presented with detailed discussions. A
brief summary is done at the final section.

\section{Lagrangian}
\idn Our lagrangian consists of a pseudoscalar meson sector ${\cal
L} ( \pi )$, a spin-1 vector and axial vector meson sector ${\cal
L} ( V, A)$, and a term of interactions with scalar particles
${\cal L}_{S}$, which comes from mass splittings in SU(3)
extension, i.e.,
\begin{equation}
{\cal L} = {\cal L}(\pi ) + {\cal L}(V, A) + {\cal L}_S~.
\end{equation}
The lagrangian for the pseudoscalar meson sector, which is a
leading Lagrangian of the ChPT, is given as
\begin{eqnarray}
\cal L(\pi) & = & \frac{f^2}{4} \langle D^\mu U^\dagger  D_\mu U
\rangle +  \frac{f^2 }{4} \langle U^\dagger \chi + \chi^\dagger U
\rangle~,
\\ D_\mu U & = & \partial_\mu U - \it i( \it v_\mu +
\it a_\mu ) U + \it i U ( \it v_\mu - \it a_\mu )~,
\end{eqnarray}
where bracket denotes a trace in flavor space, $f$ is a
pseudoscalar meson decay constant, chiral field $U = \exp( i 2\pi
/f)$ with $\pi  =  T^a \pi^a$, $T^a  =  \lambda ^a  / 2
(a=1,2,...8)$. External gauge fields are introduced via $v_\mu$
and $a_\mu$. The $\chi$ is defined by $\chi = 2B_0 (\cal S + \it i
\cal P )$. Explicit chiral symmetry
breaking due to current quark masses can be introduced by treating
those masses as if they were uniform external scalar field
$S$\cite{birse}.

Under a local $SU(N_f) \times SU(N_f)$ gauge transformation, $U
\rightarrow g_R U g_L ^\dagger $, $\chi$ and $D_\mu$ transform as
U does. The above Lagrangian is invariant provided that the
external gauge fields transform as
\begin{eqnarray}
\it v_\mu + \it a_\mu &\rightarrow& g_R ( \it v_\mu + \it a_\mu
)g_R ^\dagger - i\it \partial_\mu g_R \cdot g_R ^\dagger~,
\nonumber\\ \it v_\mu - \it a_\mu &\rightarrow& g_L ( \it v_\mu -
\it a_\mu )g_L ^\dagger -i\it \partial_\mu g_L \cdot g_L
^\dagger~, \nonumber \\ \cal S + \it i \cal P &\rightarrow&  \it
g_{\it R} ( \cal S + \it i \cal P )\it g_{\it L} ^\dagger ~ .
\end{eqnarray}
The non-linear realization of chiral symmetry is expressed in
terms of $u=\sqrt{U}$ and $h = h(u,g_R, g_L )$ defined from $ u
\rightarrow g_R u h^\dagger = hug_L ^\dagger $. In this
realization, we naturally have the following covariant quantities
\begin{eqnarray}
i\Gamma_\mu &=& \frac{i}{2} ( u^\dagger \partial_\mu u + u
\partial_\mu u^\dagger ) +\frac{1}{2} u^\dagger (v_\mu + a_\mu )u
+ \frac{1}{2}u(v_\mu - a_\mu )u^\dagger ~, \nonumber \\ i
\Delta_\mu &=& \frac{i}{2} ( u^\dagger \partial_\mu u - u
\partial_\mu u^\dagger ) +\frac{1}{2} u^\dagger ( v_\mu + a_\mu )u
- \frac{1}{2}u( v_\mu - a_\mu )u^\dagger  ~, \nonumber \\ \chi_+
&=& u^\dagger \chi u^\dagger + u \chi u~,
\end{eqnarray}
whose transformations are carried out in terms of $h$, i.e.,
$\Gamma_\mu \rightarrow h \Gamma_\mu h^\dagger -\partial_\mu h
\cdot h^\dagger $, $\Delta_\mu \rightarrow h \Delta_\mu h^\dagger
$, and $\chi_+ \rightarrow h\chi_+ h^\dagger$. With these
quantities, the Lagrangian in eq.(2) can be rewritten as
\begin{equation}
{\cal L} ( \pi )= f^2 \langle i\Delta_\mu i\Delta^\mu \rangle
+\frac{f^2}{4} \langle \chi_+ \rangle .
\end{equation}
As for the massive spin-1 mesons, we include only the mass and
kinetic terms\cite{park}
\begin{equation}
{\cal L} (V, A) =m_V ^2 \langle (V_\mu - {{i \Gamma_\mu}\over g}
)^2 \rangle + m_A ^2 \langle (A_\mu - {{i r \Delta_\mu}\over g}
)^2 \rangle - \frac{1}{2} \langle (^G V_{\mu\nu})^2 \rangle
-\frac{1}{2} \langle {( A_{\mu\nu}) }^2 \rangle
\end{equation}
with
\begin{eqnarray}
^G V_{\mu\nu} &=& \partial_\mu V_\nu - \partial_\nu V_\mu - ig[
V_\mu, V_\nu]-iG[A_\mu, A_\nu]~, \nonumber \\ A_{\mu\nu} &=&
\partial_\mu A_\nu - \partial_\nu A_\mu - ig[ V_\mu,
A_\nu]-ig[A_\mu, V_\nu]\ ~,
\end{eqnarray}
where $V_\mu = T^a V_\mu ^a $($A_\mu = T^a A_\mu ^a $) denotes
spin-1 vector (axial-vector) meson field and $g$ denotes a $V \pi
\pi$ coupling constant. The chiral transformation rules of spin-1
fields are expressed in terms of $h$
\begin{equation}
V_\mu \rightarrow hV_\mu h^\dagger - \frac{i}{g}\partial_\mu h
\cdot h^\dagger ~,~ A_\mu \rightarrow h A_\mu h^\dagger.
\end{equation}
Note that we have introduced a new form of $^G V_{\mu\nu}$. The
chiral symmetry is preserved for any value of $G$ at chiral limit
in $^G V_{\mu\nu}$, so that the value of $G$ cannot be determined
from the chiral symmetry. If $G$ is equal to $g$ as in the HGS
approach, the result may reproduce experimental data by including
other higher order terms.

The $\cal L_S$ term is introduced in the following way. It
considers effects coming from mass splittings of strange and
non-strange particles in terms of interaction Lagrangians between
scalar particles and other mesons (pseudoscalar, vector and
axial-vector mesons). In the presence of the interaction, scalar
particle field ${\cal S}$ satisfies the Klein-Gordon equation
${\cal S} \partial_{\mu} \partial^{\mu} {\cal S} + M_s {\cal S}^2
= - 2 {\cal S} J$, where ${\cal S} J$ is a source term for the
${\cal S}$ field due to the interaction. If we assume that the
kinetic term of the scalar particle is small enough to neglect
because it is massive\cite{celenza}, and integrate out the ${\cal
S}$ field from the generating function,
\begin{equation}
Z_{\cal S} = \int d {\cal S}~ exp ^{ \int [ M_S {( {\cal S} + { J
\over M_s})}^2 - { J^2 \over {M_S^2}} ]   }~,
\end{equation}
the resulting lagrangian is expressed as follows
\begin{equation}
{\cal L}_S  =  \frac{1}{M_S ^2} \langle J^2 \rangle
 =  \frac{1}{M_S ^2} \langle J^ {\prime 2} + 2 J_{vac}J^\prime
                  + J_{vac} ^2 \rangle ~,
\end{equation}
where
\begin{eqnarray}
J^ \prime & = & (J - J_{vac} ) = -  \frac{M_S}{4} s_m
\frac{2B_0}{f^2} (\pi^2 M +M\pi^2 +2\pi M\pi )+M_S j ~,\nonumber\\
j & = & s_d ( i \Delta_\mu)^2 + s_V (gV_\mu - i\Gamma_\mu )^2 +s_A
(gA_\mu - ir\Delta_\mu )^2\nonumber\\ &&+ s_r \{ i \Delta^\mu ,
gA_\mu - ir \Delta_\mu \}~, \nonumber\\ J_{vac} & = & M_S s_m B_0
M ~.
\end{eqnarray}
Here $M$ is the current quark mass matrix (we assume that masses
of u and d quarks are equal), which is given as
\begin{eqnarray}
M_a &=& 2B_0 ( \frac{D_a}{2 \sqrt{3}} \alpha + \beta ), \nonumber
\\ \frac{D_a}{2 \sqrt{3}} \alpha + \beta &=& \left\{
\begin{array}{l} \overline{m} \ for \ a=1,2,3 \\ \frac{1}{2}(
\overline{m}+m_s ) \ for \ a=4,5,6,7
\\ \frac{1}{3}( \overline{m}+2m_s ) \ for \ a=8 ~.\end{array} \right.
\end{eqnarray}
$B_0$ is a constant related to the scalar quark condensation. $j$ stands
for the interaction, on which $s_d$, $s_V$, $s_A$, $s_r$, and
$s_m$ are free parameters. It should be noted that the above
Lagrangian consists of an interaction between scalar and
pseudoscalar $(M_s \pi M ...)$, scalar and vector mesons $(M_s
j)$, and contributions from vacuum $(M_s B_0 M)$. In eq. (11), we
only consider the terms with  double and triple field products, so
that we can write down $\cal L_S$ as follows
\begin{equation}
{\cal L}_S \sim  - \frac{1}{2} (\frac{s_m}{f})^2 (\tilde M)_a ^2
(\pi^a)^2 + \frac{1}{2}s_m M_a j^{a} ,
\end{equation}
where $\tilde M_a^2 = \frac{1}{6}(2B_0\alpha)^2\delta_{8a} +
M_a^2$.
\subsection{Mixings}
\idn There are some unphysical mixings in the lagrangian, which
can be removed by field redefintion. First, let us consider a
nonet mixing. For vector bosons the nonet symmetry is good more or
less. To simplify the calculation, we use this symmetry i.e.,
$\omega$ and $\phi$ mesons are written as \cite{fayy}
\begin{equation}
\omega_\mu = \sqrt{\frac{2}{3}}\omega_{\mu1}+\sqrt{\frac{1}{3}}
V_\mu^8 ~,~ \phi_\mu
=\sqrt{\frac{1}{3}}\omega_{\mu1}-\sqrt{\frac{2}{3}}V_\mu^8.
\end{equation}
For the mixing between axial vector mesons and pion fields, we
define $A_{\mu}^{'}$ as
\begin{equation}
A_\mu =  A^\prime_\mu - \frac{r}{gf} \partial_\mu \pi .
\end{equation}
Through this field redefinition, a new term which is not
renormalizable appears in kinetic part. Therefore, in order to
keep the kinetic terms as the same as before under this field
redefinition, we also redefine vector meson field as
follows\cite{bando}
\begin{equation}
V_\mu = V^\prime_\mu -
\frac{Gr^2}{2g^2f^2}f_{abc}\pi^{b}\partial_\mu \pi^{c} .
\end{equation}
Finally, we consider the mixing between the vector meson and a
photon field. The photon field enters through the external vector
field $v_\mu = e  QA_{\mu}^{em} $, where $Q=T^3 + \frac{Y}{2}$.
The unphysical mixing related to photon field can be removed by
the following field and charge redefinitions\cite{oconell,fayy}
\begin{eqnarray}
V_{\mu} &\rightarrow& V_{\mu}^{'}+\frac{e'}{g}QA_{\mu}^{em
'}\nonumber\\ A_{\mu}^{em} &\rightarrow&
A_{\mu}^{em'}\sqrt{1-\frac{e'^2}{g^2}Q^2}\nonumber\\
e&\rightarrow&e^{'}/\sqrt{1-\frac{e'^2}{g^2}Q^2}\nonumber .
\end{eqnarray}

\subsection{Effective Lagrangian}
\idn In expanding our lagrangian, we choose only photon,
pseudoscalar meson and vector meson parts. The lagrangian is,
then, simply summarized as
\begin{eqnarray}
{\cal L}&=&\frac{1}{2}m_{Va}^2V_{\mu}V^{\mu}\nonumber\\ &
&+\frac{m_{Va}^2}{2gf_{a}^2}(1-\frac{Gr^2}{g})f_{abc}V_{\mu}^{a}\pi^{b}\partial^{\mu}\pi^{c}\nonumber\\
& &+eQf_{abc}A_{\mu}^{a}\pi^{b}\partial^{\mu}\pi^{c}\nonumber\\ &
&-\frac{1}{4}(\partial_{\mu}V_{\nu}-\partial_{\nu}V_{\mu})^2
-\frac{1}{4}(\partial_{\mu}A_{\nu}-\partial_{\nu}A_{\mu})^2\nonumber\\
&
&-\frac{e}{2g}(\partial_{\mu}V_{\nu}-\partial_{\nu}V_{\mu})(\partial^{\mu}A^{\nu}-\partial^{\nu}A^{\mu})\nonumber\\
& &-\frac{1}{2}m_{\pi
a}^2\pi^{a}\pi^{a}+\frac{1}{2}\partial_{\mu}\pi^{a}\partial^{\mu}\pi^{a},
\end{eqnarray}
where $m_{Va}^2=g^2(f_V^2 + s_ms_vM_a)$ and $V_{\mu}$ and
$A_{\mu}$ stand for redefined fields $V_{\mu}^{'}$ and
${A_{\mu}^{em}}^{'}$. In order to determine pseudoscalar meson
mass and decay constants, we exploit the following covariant
quantities \ba m_{\pi
a}^2&=&(M_{a}+(\frac{s_m}{f})^2\tilde{M_{a}}^2)/Z_{\pi a}^2~,~ f_a
= Z_{\pi a}f \nonumber\\ &with~~&  Z_{\pi a}^2 = (1+s_m
s_d\frac{M_a}{f^2})~. \ea

Mass splitting between non-strange particles and strange particles
is generated from interaction theses fields with scalar field
which is given by eq. (14).

\subsection{Comparison with ChPT}
\idn For processes with small momentum transfer, massive degree of
freedom can be integrated out leaving an effective Lagrangian of
pions and external fields, which can be compared with the
Lagrangian of ChPT. By this comparison, we can check consistency
of our Lagrangian at low energy.

To the order we consider, we can integrate out massive degree of
freedom by replacing the massive fields by their zero-th order
solutions,
\begin{eqnarray}
V_{\mu} &\rightarrow& \frac{1}{g}i\Gamma_{\mu}\nonumber\\ A_{\mu}
&\rightarrow& \frac{r}{g}i\Delta_{\mu}.
\end{eqnarray}
Then the resulting effective Lagrangian has the form
\begin{eqnarray}
\cal{L}&=&\cal{L(\pi)}\nonumber\\
&+&\frac{1}{2g^2}<(\Gamma_{\mu\nu}+r^2 G[\Delta_{\mu},
\Delta_{\nu}])^2>+\frac{r^2}{2g^2}<{\Delta_{\mu\nu}^2}>
\end{eqnarray}
with
\begin{eqnarray}
\Delta^{\mu\nu}&=&\partial^{\mu}\Delta^{\nu}-\partial^{\nu}\Delta^{\mu}+
[\Gamma^{\mu}, \Delta^{\nu}]-[\Gamma^{\nu},
\Delta^{\mu}]\nonumber\\
 &=&-\frac{i}{2}(\xi^{\dagger}F^{\mu\nu}_{R}\xi-\xi F^{\mu\nu}_{L}\xi^{\dagger}),\nonumber\\
\Gamma^{\mu\nu}&=&\partial^{\mu}\Gamma^{\nu}-\partial^{\nu}\Gamma^{\mu}+[\Gamma^{\mu},\Gamma^{\nu}]\nonumber\\
&=&-[\Delta^{\mu},\Delta^{\nu}]-\frac{i}{2}(\xi^{\dagger}F^{\mu\nu}_{R}\xi-\xi
F^{\mu\nu}_{L}\xi^{\dagger})
\end{eqnarray}
and
\begin{eqnarray}
F^{\mu\nu}_{L,R}&=&\partial^{\mu}(v^{\nu}\mp
a^{\nu})-\partial^{\nu}(v^{\mu}\mp a^{\mu})\nonumber\\
&-&i[v^{\mu}\mp a^{\mu}, v^{\nu}\mp a^{\nu}].
\end{eqnarray}
For easy comparison, we list the contribution to the coefficient
of ${\cal L}_{4}$:
\begin{eqnarray}
L^{V}_{1}&=&\frac{1}{32g^2}(1-Gr^2)^2, L^{V}_{2}=2L^{V}_{1},
L^{V}_{3}=-6L^{V}_{1},\nonumber\\
L^{V}_{9}&=&\frac{1}{4g^2}(1-Gr^2)^2, L^{V}_{10}=-\frac{1}{4g^2},
H^{V}_{1}=\frac{1}{2}L^{V}_{10},\nonumber\\
L^{A}_{10}&=&\frac{r^2}{4g^2}, H^{A}_{1}=-\frac{1}{2}L^{A}_{10},
\end{eqnarray}
which are equivalent with Ecker et.al.'s expressions (eqs.(5.6)
and (5.9) of ref.\cite{ecker}) with the following correspondence
of parameters between the two formula,
\begin{eqnarray}
g&\longleftrightarrow&\frac{M_{V}}{F_{V}},\nonumber\\
Gr^2&\longleftrightarrow&\frac{F_{V}-2G_{V}}{F_{V}},\nonumber\\
r&\longleftrightarrow&\frac{F_{A}}{F_{V}}\frac{M_{V}}{M_{A}},
\end{eqnarray}
where $F_{V},G_{V},F_{A},M_{V}\simeq m_{V}$  and $M_{A}\simeq
m_{A}$ are the parameters introduced by them.

\section{Pion and Kaon Electromagnetic Form Factor}
The pion form-factor in the time-like region is dominated by the
$\rho$-meson resonance. Likewise to the pion the kaon form-factor
is influenced mainly by the $\phi$-meson. But the contribution of
$\rho$-$\omega$ meson mixing is also important. With the effective
lagrangian in section 2, we improve the analysis of both
form-factors. The pseudoscalar meson loops are also taken into
account.

\subsection{The $\rho$ meson self energy}
From the effective lagrangian, V-$\pi$ interaction term (the 2nd
term in eq.(18)) generates a vector current of pion as \ba
J_{had}^{\mu}=i(\pi^{+}\partial^{\mu}\pi^{-}-\pi^{-}\partial^{\mu}\pi^{+})
. \ea This coupling to the $\rho$-meson field produces the self
energy as shown in Fig 1, which is calculated as
\be
-i\Pi^{\mu\nu}=g_{\rho\pi\pi}^2\int\frac{d^{4}p}{(2\pi)^4}\frac{(2p-q)^{\mu}(2p-q)^{\nu}}
{(p^2-m_{\pi}^2+i\epsilon)((p-q)^2-m_{\pi}^2+i\epsilon)}~. \ee The
$\rho$ meson coupling to a conserved current implies that the self
energy is transverse, i.e.,
\be
q_{\mu}\Pi^{\mu\nu}(q)=q_{\nu}\Pi^{\mu\nu}(q)=0 . \ee This
property, combined with Lorentz invariance, uniquely determines
the tensor structure of the self energy
\be
\Pi^{\mu\nu}=(-g^{\mu\nu}+\frac{q^{\mu}q^{\nu}}{q^2})\Pi(q^2) .
\ee The full propagator of $\rho$ meson is then given as \ba
D^{\mu \nu}=\frac{1}{q^2-\dot{m}_{\rho}^{2}-\Pi_{\rho}}
(-g^{\mu\nu}+\frac{q^{\mu}q^{\nu}}{q^2})+
\frac{1}{\dot{m}_{\rho}^{2}}\frac{q^{\mu}q^{\nu}}{q^2}~. \ea
Here,
 the bare $\rho$ meson mass, $\dot{m_{\rho}}$, is introduced so that
its physical mass is given by
\be
m_{\rho {\rm
P}}^{2}=\dot{m}_{\rho}^2+\Re[\Pi_{\rho}(q^2=m_{\rho}^2)] . \ee
Since the full propagator of $\rho$ meson is given as eq.(30),
the $\rho\rightarrow\pi\pi$ decay width at resonance is given as
\be
\Gamma_{\rho\rightarrow\pi\pi}=-\Im\Pi_{\rho}(q^2=m_{\rho}^2)/m_{\rho}
. \ee
\subsection{Regularization}
To calculate the self energy of Fig.1, we use a Pauli-Villars
regularization method\cite{herrmann}. The regularized self energy
is given as
\be
\Pi^{\mu\nu}(q)=\tilde{\Pi}^{\mu\nu}(q,m_{\pi})-\sum
B_{i}\tilde{\Pi}^{\mu\nu} (q,\Lambda_{i})~. \ee
Once the
fictitious higher masses $\Lambda_{i}$ are fixed, the coefficients
$B_{i}$ are determined by requiring that the self energy be
finite. Since this term, in its unregularized form, is
quadratically divergent, we need two subtractions and obtain
\be
B_{1}=\frac{\Lambda_{2}^2-m_{\pi}^2}{\Lambda_{2}^2-\Lambda_{1}^2}~,~
B_{2}=\frac{\Lambda_{1}^2-m_{\pi}^2}{\Lambda_{1}^2-\Lambda_{2}^2}
. \ee
Then, from the conditions that $\Lambda_{2}$ goes to infinite and
$\Lambda_{1}$ is fixed to 1 GeV, regularized self energy is
obtained as follows \ba
\Re[\Pi]&=&-\frac{g_{\rho\pi\pi}^2}{24\pi^2}q^2\left({\cal
G}(q,m_{\pi})-{\cal G}(q,\Lambda_{1})
+4(\Lambda_{1}^2-m_{\pi}^2)/q^2+\rm{ln}(\frac{\Lambda_{1}}{m_{\pi}})\right)~,
\nonumber\\
\Im[\Pi]&=&-\frac{g_{\rho\pi\pi}^2}{48\pi}q^2\left((1-\frac{4m_{\pi}^2}{q^2})^{3/2}\Theta(q^2-4m_{\pi}^2)
-(1-\frac{4\Lambda_{1}^2}{q^2})^{3/2}\Theta(q^2-4\Lambda_{1}^2)\right),\nonumber\\
\ea where \ba {\cal
G}&=&(\frac{4m_{\pi}^2}{q^2}-1)^{3/2}\arctan(\sqrt{\frac{4m_{\pi}^2}{q^2}-1}^{~-1})
,\hspace{0.5cm}0<q^2<4m_{\pi}^2\nonumber\\
&=&-\frac{1}{2}(1-\frac{4m_{\pi}^2}{q^2})^{3/2}\ln\frac{\sqrt{\frac{4m_{\pi}^2}{q^2}-1}+1}
{\sqrt{\frac{4m_{\pi}^2}{q^2}-1}-1}
,\hspace{0.5cm}4m_{\pi}^2<q^2,\hspace{0.5cm}q^2<0\nonumber . \ea

\subsection{Renormalization}
We consider a $\rho$ meson renormalization process. Expanding the
transverse part of the $\rho$ meson  propagator around the
physical mass $m_{\rho}$, we have
\be
\frac{1}{q^2-\dot{m}_{\rho}^2-\Pi_{\rho}(q^2)}\approx\frac{Z}{(q^2-m_{\rho}^2)-iZ\Im\Pi_{\rho}(q^{2})},
\ee where $Z$ is a renormalization constant given by
\be
Z=(1-\frac{d}{dq^2}\Re\Pi_{\rho}(q^2)\mid_{(q^2=m_{\rho}^2)})^{-1}.
\ee Here we introduce a bare coupling constant
$\dot{g}_{\rho\pi\pi}$ which is related to the physical coupling
constant $g_{\rho\pi\pi}$ by $\dot{g}_{\rho\pi\pi}=Z^{1/2}
g_{\rho\pi\pi}$. With the condition $Z=1$, which means that the
real part of the self energy is completely absorbed by shifting
the
 bare mass $\dot{m}_{\rho}$ to its physical value $m_{\rho}$ with
 coupling constant $g_{\rho\pi\pi}$ remained unaltered, we
 can determine a new condition
\be
\frac{d}{dq^2}\Re\Pi_{\rho}\mid_{(q^2=m_{\rho}^2)}=0 . \ee In
order to satisfy this condition, we need to add an arbitrary
constant term $c_{\pi}q^2$ in real part of the self energy, where
$c_{\pi}$ is given by
\begin{eqnarray}
c_{\pi}=\frac{5.774(g-Gr^2)^2}{g^4} .
\end{eqnarray}

\subsection{Pion Electromagnetic Form-factor}
The electromagnetic pion form-factor is defined by the following
matrix element
\begin{eqnarray}
\langle \pi^\pm(k')|J_\mu^{em}(0)|\pi^\pm(k)\rangle = \pm
(k+k')_\mu F_\pi (q^2).
\end{eqnarray}
The leading term of $F_{\pi}(q^2)$ obtained from ${\cal
L}_{\gamma\pi}$, ${\cal L}_{\gamma V}$, (3rd and 5th terms,
respectively) in the lagrangian, is expressed in the following way
\be
F_{\pi}^{(o)}(q^2)=1-\frac{g_{\rho\pi\pi}}{g}\frac{q^2}{q^2-m_{\rho}^{2}+im_{\rho}\Gamma_{\rho}},
\ee
where $g$ is a bare coupling constant which does not consider
the loop effect. $m_{\rho}$ and $\dot{m}_{\rho}$ are means of
$\rho$ meson and bare $\rho$ meson, respectively. The
relation of both masses is given by eq.(31).
 Introducing the $\rho\pi\pi$ self energy in Fig.
2, we obtained
\begin{eqnarray}
F_{\pi}(q^2)&=&1-\frac{g_{\rho\pi\pi}}{g}\frac{q^2}{q^2-\dot{m}_{\rho}^{2}-\Pi_{\rho}}+
\frac{\Pi_{\rho}}{q^2-\dot{m}_{\rho}^{2}-\Pi_{\rho}}\nonumber\\
&=&1-\frac{g_{\rho\pi\pi}}{g(q^{2})}\frac{q^2}{q^2-\dot{m}_{\rho}^{2}-\Pi_{\rho}}
.
\end{eqnarray}
Note that not only the $\rho$ meson propagator, but also
$\gamma-\rho$ coupling  is modified by the pion loop as shown in
Fig. 3 as follows
\begin{eqnarray}
-\frac{eq^2}{g(q^2)}=-\frac{eq^2}{g}+\frac{e\Pi_{\rho}}{g_{\rho\pi\pi}}.
\end{eqnarray}
The constant $g$ determined from the experimental $\rho\rightarrow
e^+ e^-$ decay width should be compared with
$\Re[g(q^2)]_{q^2=m_{\rho}^2}$. Using
$\Re[g(q^2)]_{q^2=m_{\rho}^2}$ and experimental results of
$\Gamma_{\rho\rightarrow\pi\pi}$, we find $g$ and $\beta=Gr^2$
values. When $g$ is 5.36 and $\beta$ is 0.32, $g_{\rho\pi\pi}$
goes to 6.037. Under universality ($g_{\rho
\pi\pi}=g_{\rho\gamma}$) used in VMD model, the prediction of the
pion form factor is underestimated compared to the experimental
values. Brown et.al \cite{brown} allows its violation i.e.
$\beta={g_{\rho\pi\pi}/g_{\rho\gamma}}=1.2$ by considering the
intrinsic size due to the vector meson. In the papers, for example
of Brown\cite{brown} and Klingl\cite{weise}, this contribution is
attributed to those of vector mesons, which are incorporated by
the gauge fields in the hidden gauge symmetry approach, while in
this paper these fields are introduced explicitly using the
non-linear realization of the chiral symmetry exploited originally
by Weinberg.

Using eqs. (31) and (32), we also find physical mass and decay
width:
\begin{equation}
m_{\rho P} \approx 771\MeV ~,~ \Gamma_{\rho\rightarrow\pi\pi}
\approx 150\MeV ~,~ \dot{m_{\rho}} \approx 808\MeV ~.
\end{equation}
Finally, the inclusion of $\rho-\omega$ mixing turned out to give
another factor to eq.(42) in the following way
\begin{equation}
F_{\pi}(q^2)=(1-\frac{g_{\rho\pi\pi}}{g(q^{2})}\frac{q^2}{q^2-\dot{m}_{\rho}^{2}-\Pi_{\rho}})
(1+\frac{g({q^{2}})}{g_{\omega}}\frac{z_{\rho\omega}}{q^2-m_{\omega}^{2}-im_{\omega}\Gamma_{\omega}})
.
\end{equation}
The $\omega$ meson width $\Gamma_{\omega}$=8.4MeV is used and the
mixing parameter $z_{\rho\omega}=-4.52\times 10^{-3}\rm{GeV}^{2}$
from Ref. \cite{weise} is also exploited. The corresponding
optimal result for $F_{\pi} (q^2)$ compared with experimental
data\cite{ba} is shown in Fig. 4. Dashed line plots eq.(41), which
corresponds to VMD model prediction. Dotted line plots eq.(42)
i.e. one loop correction is included, while solid line represents
eq.(45) in which $\rho-\omega$ mixing contribution is taken into
account.

The form factor in the space like region ($q^{2}<0$) is also given
in Fig. 5. Our approach gives a good agreement with experimental
result~\cite{amen84}. The squared pion charged radius becomes

\begin{eqnarray}
<{r_{\pi}^{2}}>&=& 6\frac{dF_{\pi}}{dq^{2}}|_{q^{2}=0}\nonumber\\
&=&
\frac{6}{\dot{m}_{\rho}^2}(\frac{g_{\rho\pi^{+}\pi^{-}}}{g}-c_{\pi}
+\frac{g_{\rho\pi^{+}\pi^{-}}^{2}}{24\pi^{2}}\ln(\frac{\Lambda_{1}}{m_{\pi}}))\nonumber\\
&=&0.447fm^{2}.
\end{eqnarray}
Using the constants determined before, eq. (46) yields a good
agreement with experimental value $<{r_{\pi}^{2}}>=(0.44\pm
0.01)fm^{2}$. From KSRF relation
$\dot{m}_{\rho}^2=2g_{\rho\pi\pi}^2f_{\pi}^2$, mean square radius
of pion becomes
\begin{equation}
<{r_{\pi}^2}>=(\frac{1}{4\pi
f_{\pi}})^{2}\ln(\frac{\Lambda_{1}^2}{m_{\pi}^2}) + constant~ .
\end{equation}
Therefore, in the chiral limit ($m_\pi\rightarrow0$), the pion
radius diverges logarithmically,which is consistent with
ChPT\cite{weise}.
\subsection{Kaon Electromagnetic Form-factor}
The electromagnetic form-factor of charged kaon is also defined by
\begin{eqnarray}
\langle {\rm K}^\pm(k')|J_\mu^{em}(0)|{\rm K}^\pm(k)\rangle = \pm
(k+k')_\mu F_{K}(q^2).
\end{eqnarray}
The leading behavior of $F_{K}(q^2)$ is obtained just by
transcribing the previous formalism developed for $F_{\pi}(q^2)$
in eq. (42) and replacing the $\rho$ meson by the $\phi$ meson and
the pion loop by the kaon loop. It leads to yield the following result
obtained
\begin{eqnarray}
F_{K}(q^2)&=&1+\frac{\sqrt{2}}{3}\frac{g_{\phi
K^{+}K^{-}}}{g}\frac{q^2}{q^2-\dot{m}_{\phi}^{2}-\Pi_{\phi}}+
\frac{\Pi_{\phi\rightarrow
K^{+}K^{-}}}{q^2-\dot{m}_{\phi}^{2}-\Pi_{\phi}}\nonumber\\
&=&1+\frac{\sqrt{2}}{3}\frac{g_{\phi
K^{+}K^{-}}}{g(q^{2})}\frac{q^2}{q^2-\dot{m}_{\phi}^{2}-\Pi_{\phi}}~,
\end{eqnarray}
where the $\phi$ meson self energy has the contributions not only
from $ K^+K^-$ but also from $K^{o}_{L}K^{o}_{S}$, i.e.,
\begin{eqnarray}
\Pi_{\phi}=\Pi_{\phi\rightarrow K^+K^-} +\Pi_{\phi\rightarrow
K^{o}_{L}K^{o}_{S}}~.
\end{eqnarray}

The photon coupling of the $\phi$ meson is also modified by means
of the charged kaon loop including the renormalization
\begin{eqnarray}
\frac{1}{g_{\phi}(q^2)}=\frac{1}{g}+\frac{3}{\sqrt{2}}\frac{\Pi_{\phi\rightarrow
K^{+}K^{-}}}{g_{\phi K^+K^-}q^2} .
\end{eqnarray}
Considering the additional contributions of both $\rho$ meson and
$\omega$ meson. We obtain the final form of the charged kaon
form-factor \ba F_{K}(q^2)&=&1+\frac{\sqrt{2}}{3}\frac{g_{\phi
K^{+}K^{-}}}
{g_{\phi}(q^{2})}\frac{q^2}{q^2-\dot{m}_{\phi}^{2}-\Pi_{\phi}}\nonumber\\
& &-\frac{g_{\rho
K^{+}K^{-}}}{g_{\rho}(q^{2})}\frac{q^2}{q^2-\dot{m}_{\rho}^{2}-\Pi_{\rho}}
-\frac{4}{3}
\frac{g_{\omega
K^{+}K^{-}}}{g_{\omega}(q^{2})}\frac{q^2}{q^2-\dot{m}_{\omega}^{2}-\Pi_{\omega}}
.
\end{eqnarray}
Fig.6 shows the charged kaon form-factor at the time-like
region compared with experimental data \cite{iva}. Here the
coupling constant ${\rm Re}[g_{\phi}(q^2)]$ is approximately 6.5.
And physical $\phi$ meson mass and $\phi\rightarrow K^+ K^-$ ,
$\phi\rightarrow K_{S}^0 K_{L}^0$ decay width are given by
\begin{eqnarray}
& &m_{\phi P}\approx 1019 {\rm MeV} ~,~ \dot{m}_{\phi}\approx 940
{\rm MeV}\nonumber\\ & &\Gamma_{\phi\rightarrow K^+ K^-}\approx
2.32 {\rm MeV} ~,~ \Gamma_{\phi\rightarrow K_{S}^0 K_{L}^0}\approx
1.517 {\rm MeV} \nonumber .
\end{eqnarray}
The kaon form factor at space like region $q^2<0$ is shown in fig. 7.
The final inclusion of the $\rho$ and $\omega$ meson contributions
reproduce successfully experimental results\cite{amen}.
The calculation of the mean square radius of
the charged kaon is also successfully performed with the $\rho$ and
$\omega$ meson contribution
\begin{eqnarray}
<{r_{{K}^{\pm}}}>&=& 6\frac{dF_{K}}{dq^{2}}|_{q^{2}=0}\nonumber\\
&=&\frac{6}{\dot{m}_{\rho}^2}(\frac{g_{\rho
K^{+}K^{-}}}{g}-c_{\rho K} +\frac{g_{\rho
K^{+}K^{-}}^{2}}{24\pi^{2}}\ln(\frac{\Lambda_{1}}{m_{K}}))\nonumber\\
&+&\frac{6}{\dot{m}_{\omega }^2}(\frac{4}{3}\frac{g_{\omega
K^{+}K^{-}}}{g}-c_{\omega K} +\frac{g_{\omega
K^{+}K^{-}}^{2}}{24\pi^{2}}\ln(\frac{\Lambda_{1}}{m_{K}}))\nonumber\\
&-&\frac{6}{\dot{m}_{\phi}^2}(\frac{\sqrt{2}}
{3}\frac{g_{\phi
K^{+}K^{-}}}{g}+c_{\phi K} -\frac{g_{\phi
K^{+}K^{-}}^{2}}{24\pi^{2}}\ln(\frac{\Lambda_{1}}{m_{K}}))\nonumber\\
&=&0.332fm^{2}~,
\end{eqnarray}
where $c_{\rho K}, c_{\omega K}$,and $c_{\phi K}$ determined from
the renormalization condition eq.(39) are $c_{\rho K}=0.0308$,
$c_{\omega K}=0.0175$ and $c_{\phi K}=0.171$.
This value agrees well with the experimental value \cite{amen}
\begin{equation}
<{r_{K_\pm}^2}>=(0.34\pm0.05)fm^2 .
\end{equation}
Likewise to the case of pion, in the chiral limit,
the mean squared radius of charged kaon also diverges logarithmically.

Finally, the neutral kaon mean square radius is explored. It is
obtained from eq.(53) replacing the $K^{+}K^{-}$ coupling by
$K^{o}_{L}K^{o}_{S}$. The sign for the $\rho$ meson contribution
is changed by $g_{\rho K^{o}_{L}K^{o}_{S}}=-g_{\rho
K^{+}K^{-}}=0.201$, while $\phi$ meson contribution is remained by
$g_{\phi K^{o}_{L}K^{o}_{S}}=g_{\phi K^{+}K^{-}}=-4.72$. But, from
nonet mixing, $\omega$ meson contribution in the case of neutral
kaon is different from that of charged kaon as follows,
\begin{eqnarray}
g_{\omega K^{+}K^{-}}&=&\frac{(m_{\phi}^2+m_{\omega}^2)}{4gf_{K}^2}
(1-\frac{\beta_{\omega K}}{g})=1.51\nonumber\\
g_{\omega K^{o}_{L}K^{o}_{S}}&=&\frac{(m_{\phi}^2-m_{\omega}^2)}{4gf_{K}^2}
(1-\frac{\beta_{\omega K}}{g})=0.17~.
\end{eqnarray}
These constants play important role of understanding the mean
square radii of both neutral and charged kaon.
Final form of mean
square radius of neutral kaon is given as
\begin{eqnarray}
<{r_{{K}^{o}}}>&=& 6\frac{dF_{K}}{dq^{2}}|_{q^{2}=0}\nonumber\\
&=&-\frac{6}{\dot{m}_{\rho}^2}
(\frac{g_{\rho
K^{o}_{L}K^{o}_{S}}}{g}+c_{\rho K^{o}}-\frac{g_{\rho
K^{o}_{L}K^{o}_{S}}}{24\pi^{2}}\ln(\frac{\Lambda_{1}}{m_{K}}))\nonumber\\
&+&\frac{6}{\dot{m}_{\omega }^2}(\frac{4}{3}\frac{g_{\omega
K^{o}_{L}K^{o}_{S}}}{g}-c_{\omega K^{o}} +\frac{g_{\omega
K^{o}_{L}K^{o}_{S}}}{24\pi^{2}}\ln(\frac{\Lambda_{1}}{m_{K}}))\nonumber\\
&-&\frac{6}{\dot{m}_{\phi}^2}(\frac{\sqrt{2}}
{3}\frac{g_{\phi
K^{o}_{L}K^{o}_{S}}}{g}+c_{\phi K^{o}} -\frac{g_{\phi
K^{o}_{L}K^{o}_{S}}^{2}}{24\pi^{2}}\ln(\frac{\Lambda_{1}}{m_{K}}))\nonumber\\
&=&-0.0549fm^{2}~.
\end{eqnarray}
The different role of intermediate $\omega$ meson contribution to
$K^{o}$ perfectly reproduces empirical value\cite{molzon}
\begin{equation}
<r^{2}_{K^{O}}>=(-0.054\pm0.0026)fm^{2}~.
\end{equation}

\section{Conclusion}
We extended a chiral effective lagrangian by including the vector
and the axial-vector mesons as well as pions to $SU_{R}(3)\otimes
SU_{L}(3)$. The meson fields are introduced through the non-linear
realization of chiral symmetry, which provides an easy way of
imposing consistency with the ChPT. In order to have mass
splitting of strange and non-strange particles, the interactions
between scalar mesons and each meson i.e, vector, axial-vector,
and pseudoscalar mesons are taken into account.

For phenomenological side of this lagrangian, pion and kaon
electromagnetic form-factors and some related decays are
calculated. In the process of calculating, our effective
Lagrangian is shown to give a good agreement with experimental
data without considering the effects from the higher orders in
other effective theories.
\vskip0.5cm
This work was supported in
part by the Korea Research Foundation.

\newpage
\begin{figure}
\begin{center}
\epsfysize3cm \leavevmode\epsfbox{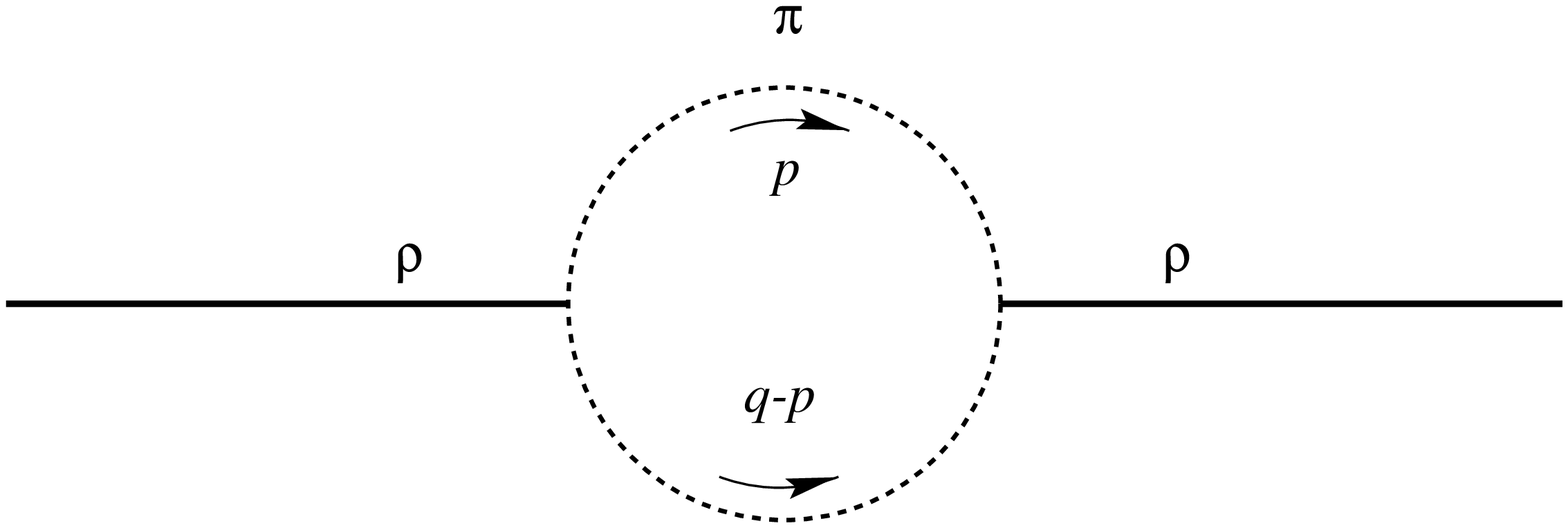} \caption{$\rho$ meson
self energy}
\end{center}
\end{figure}

\begin{figure}
 \begin{center}
 \epsfysize3cm \leavevmode\epsfbox{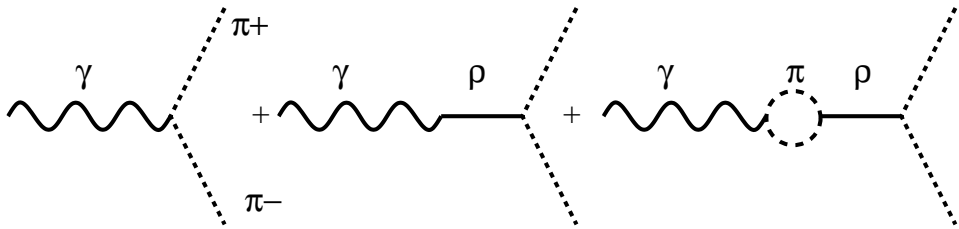}
 \caption{Feynmann Diagram of Pion Form-factor with self energy}
 \end{center}
 \end{figure}

 \begin{figure}
  \begin{center}
  \epsfysize3cm \leavevmode\epsfbox{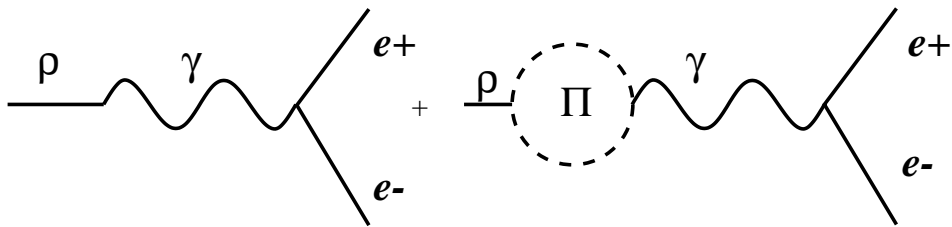}
  \caption{$\rho$ meson decay}
   \end{center}
   \end{figure}

   \begin{figure}
   \begin{center}
   \epsfysize10cm \leavevmode\epsfbox{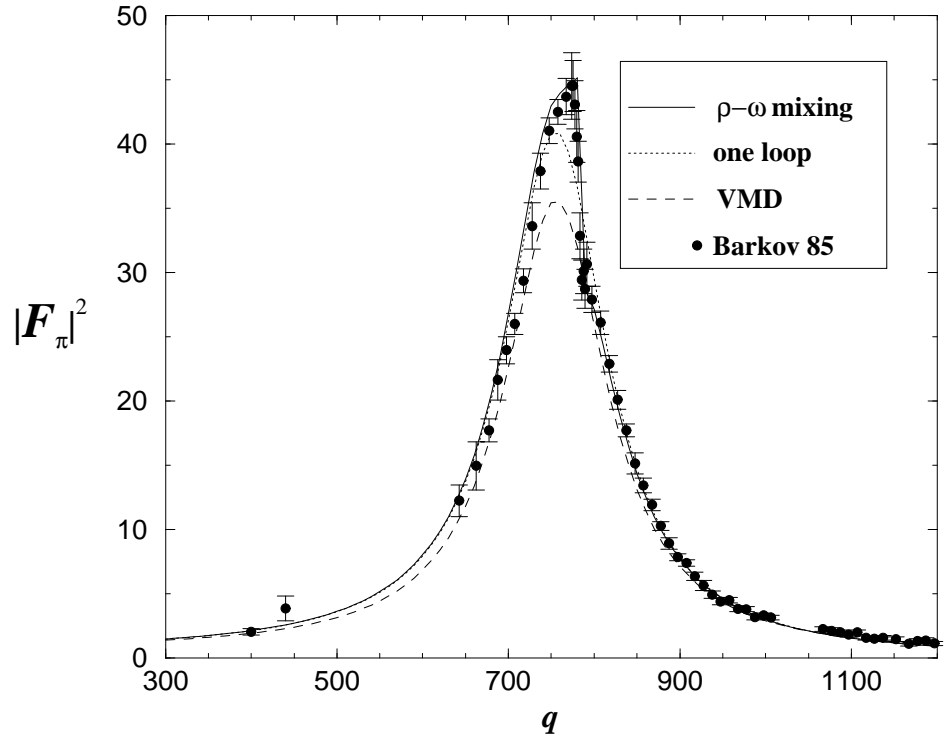}
   \caption{Pion Electromagnetic Form-Factor in time like region:
   solid, dotted and dashed lines represent eqs.(45), (42) and (41), respectively.
   Here $q$ means ${\sqrt { q^2}}$}
   \end{center}
   \end{figure}

   \begin{figure}
   \begin{center}
   \epsfysize10cm \leavevmode\epsfbox{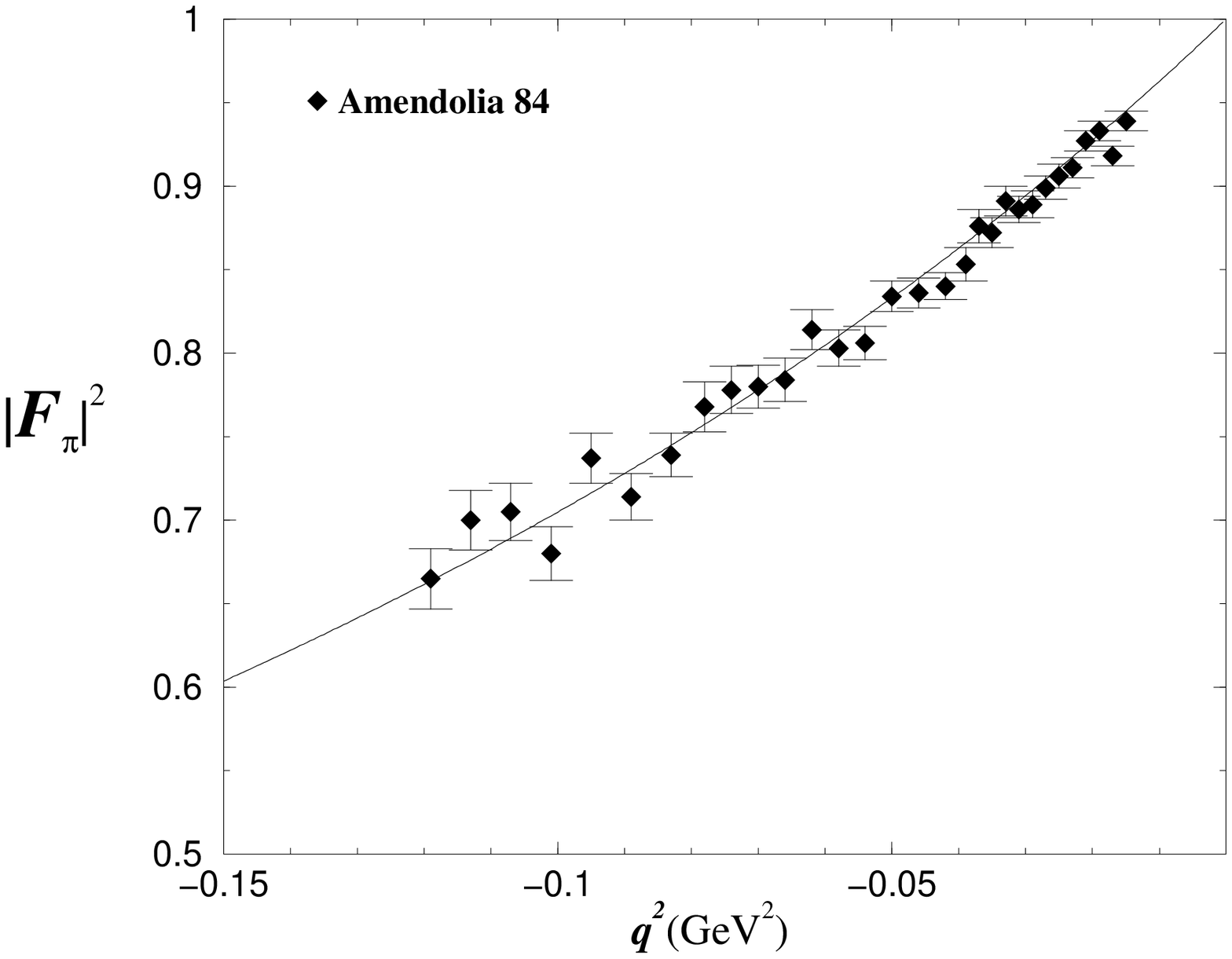}
   \caption{Pion Electromagnetic Form-Factor in space like region}
   \end{center}
   \end{figure}

   \begin{figure}
   \begin{center}
   \epsfysize9cm \leavevmode\epsfbox{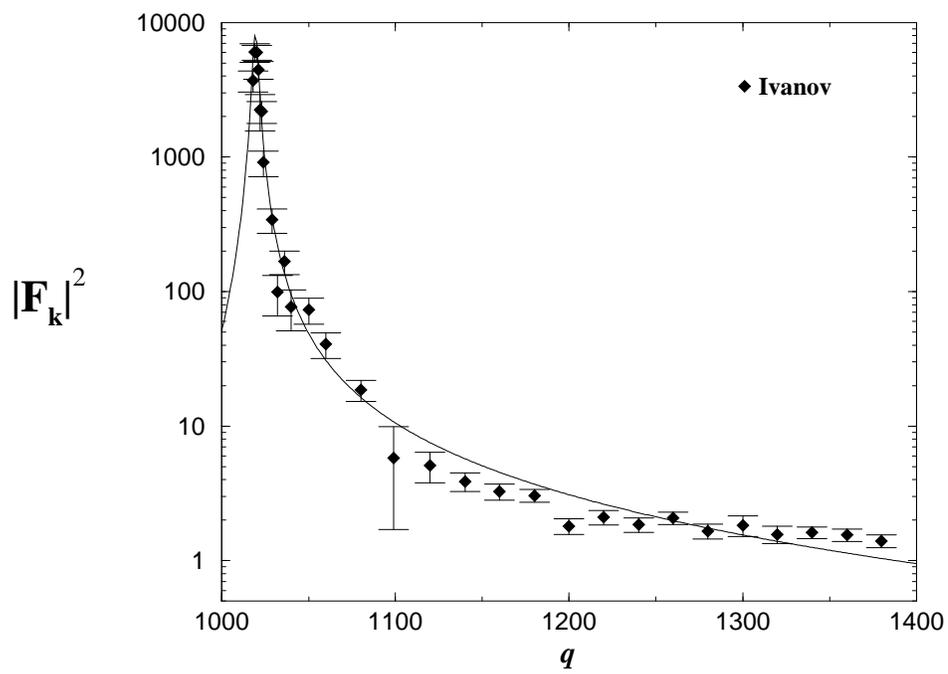}
   \caption{Kaon Electromagnetic Form-Factor in time like region :
   Here $q$ means ${\sqrt { q^2}}$}
   \end{center}
   \end{figure}

   \begin{figure}
    \begin{center}
    \epsfysize9cm \leavevmode\epsfbox{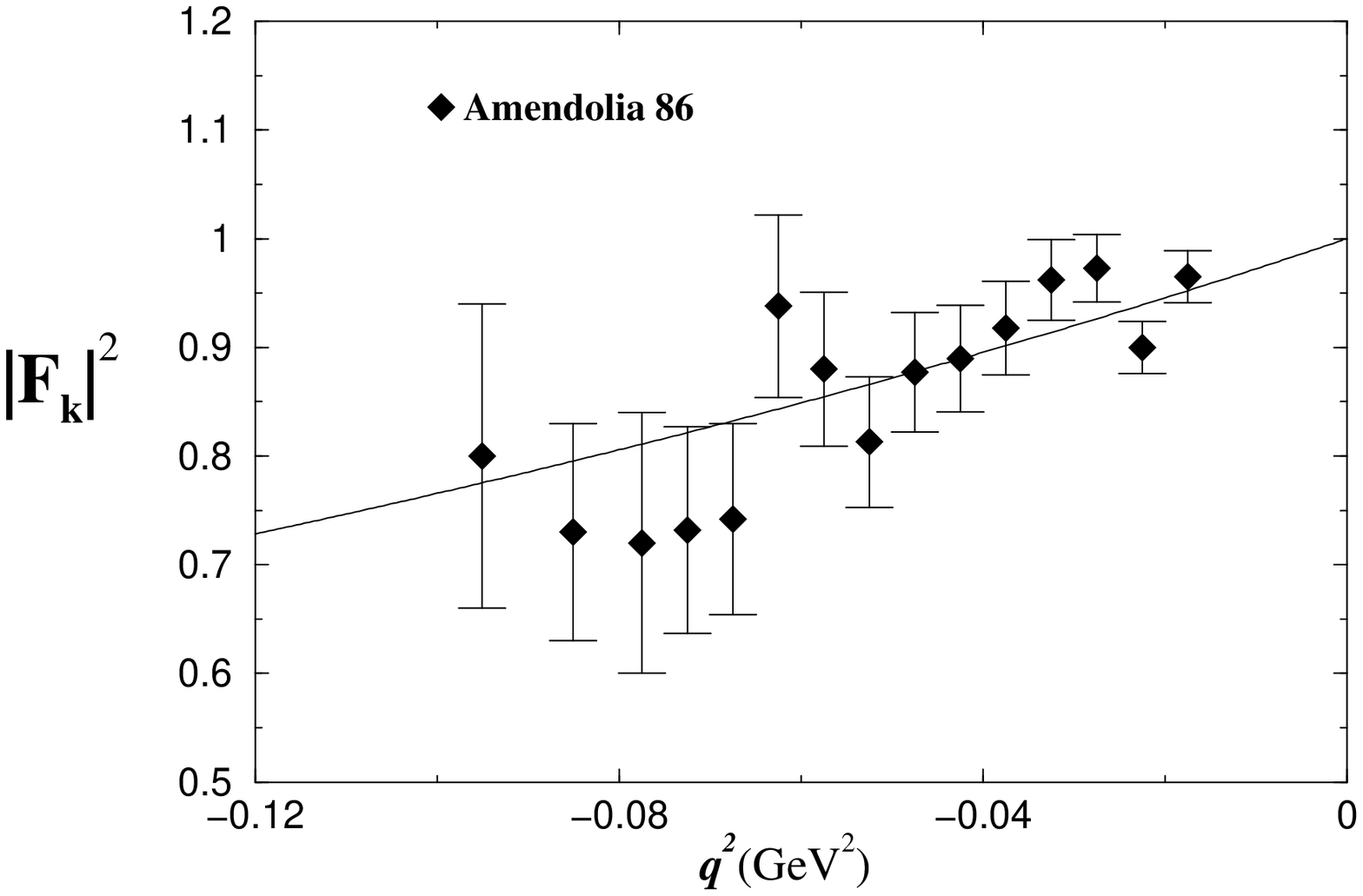}
    \caption{Kaon Electromagnetic Form-Factor in space like region}
    \end{center}
    \end{figure}

\end{document}